\documentclass[preprint,tightenlines,superscriptaddress,showpacs,byrevtex]{revtex4}
\usepackage{times}
\usepackage{amsmath}
\usepackage{graphicx}
\usepackage{epsfig}
\usepackage{threeparttable}
\usepackage{dcolumn}
\usepackage{bm}
\usepackage{multirow}
\usepackage{url}
\usepackage[colorlinks,linkcolor=red,anchorcolor=green,citecolor=blue]{hyperref}
\usepackage{array}
\usepackage{booktabs}

\newcommand{\BR}{{\cal B}}

\newcommand{\jpsi}{J/\psi}

\newcommand{\EE}{e^+e^-}

\newcommand{\LL}{\ell^+\ell^-}
\newcommand{\pp}{\pi^+\pi^-}

\newcommand{\beq}{\begin{equation}}
\newcommand{\eeq}{\end{equation}}
\newcommand{\bitm}{\begin{itemize}}
\newcommand{\eitm}{\end{itemize}}

\begin{document}


\title{
\boldmath Search for light tetraquark states in $\Upsilon(1S)$ and $\Upsilon(2S)$ decays}


\noaffiliation
\affiliation{University of the Basque Country UPV/EHU, 48080 Bilbao}
\affiliation{Beihang University, Beijing 100191}
\affiliation{Budker Institute of Nuclear Physics SB RAS, Novosibirsk 630090}
\affiliation{Faculty of Mathematics and Physics, Charles University, 121 16 Prague}
\affiliation{Chonnam National University, Kwangju 660-701}
\affiliation{University of Cincinnati, Cincinnati, Ohio 45221}
\affiliation{Deutsches Elektronen--Synchrotron, 22607 Hamburg}
\affiliation{University of Florida, Gainesville, Florida 32611}
\affiliation{Department of Physics, Fu Jen Catholic University, Taipei 24205}
\affiliation{Justus-Liebig-Universit\"at Gie\ss{}en, 35392 Gie\ss{}en}
\affiliation{Gifu University, Gifu 501-1193}
\affiliation{SOKENDAI (The Graduate University for Advanced Studies), Hayama 240-0193}
\affiliation{Gyeongsang National University, Chinju 660-701}
\affiliation{Hanyang University, Seoul 133-791}
\affiliation{University of Hawaii, Honolulu, Hawaii 96822}
\affiliation{High Energy Accelerator Research Organization (KEK), Tsukuba 305-0801}
\affiliation{J-PARC Branch, KEK Theory Center, High Energy Accelerator Research Organization (KEK), Tsukuba 305-0801}
\affiliation{IKERBASQUE, Basque Foundation for Science, 48013 Bilbao}
\affiliation{Indian Institute of Science Education and Research Mohali, SAS Nagar, 140306}
\affiliation{Indian Institute of Technology Bhubaneswar, Satya Nagar 751007}
\affiliation{Indian Institute of Technology Guwahati, Assam 781039}
\affiliation{Indian Institute of Technology Hyderabad, Telangana 502285}
\affiliation{Indian Institute of Technology Madras, Chennai 600036}
\affiliation{Indiana University, Bloomington, Indiana 47408}
\affiliation{Institute of High Energy Physics, Chinese Academy of Sciences, Beijing 100049}
\affiliation{Institute of High Energy Physics, Vienna 1050}
\affiliation{Institute for High Energy Physics, Protvino 142281}
\affiliation{University of Mississippi, University, Mississippi 38677}
\affiliation{INFN - Sezione di Napoli, 80126 Napoli}
\affiliation{INFN - Sezione di Torino, 10125 Torino}
\affiliation{Advanced Science Research Center, Japan Atomic Energy Agency, Naka 319-1195}
\affiliation{J. Stefan Institute, 1000 Ljubljana}
\affiliation{Kanagawa University, Yokohama 221-8686}
\affiliation{Institut f\"ur Experimentelle Kernphysik, Karlsruher Institut f\"ur Technologie, 76131 Karlsruhe}
\affiliation{Kennesaw State University, Kennesaw, Georgia 30144}
\affiliation{King Abdulaziz City for Science and Technology, Riyadh 11442}
\affiliation{Department of Physics, Faculty of Science, King Abdulaziz University, Jeddah 21589}
\affiliation{Korea Institute of Science and Technology Information, Daejeon 305-806}
\affiliation{Korea University, Seoul 136-713}
\affiliation{Kyoto University, Kyoto 606-8502}
\affiliation{Kyungpook National University, Daegu 702-701}
\affiliation{\'Ecole Polytechnique F\'ed\'erale de Lausanne (EPFL), Lausanne 1015}
\affiliation{P.N. Lebedev Physical Institute of the Russian Academy of Sciences, Moscow 119991}
\affiliation{Faculty of Mathematics and Physics, University of Ljubljana, 1000 Ljubljana}
\affiliation{Ludwig Maximilians University, 80539 Munich}
\affiliation{University of Malaya, 50603 Kuala Lumpur}
\affiliation{University of Maribor, 2000 Maribor}
\affiliation{Max-Planck-Institut f\"ur Physik, 80805 M\"unchen}
\affiliation{School of Physics, University of Melbourne, Victoria 3010}
\affiliation{University of Miyazaki, Miyazaki 889-2192}
\affiliation{Moscow Physical Engineering Institute, Moscow 115409}
\affiliation{Moscow Institute of Physics and Technology, Moscow Region 141700}
\affiliation{Graduate School of Science, Nagoya University, Nagoya 464-8602}
\affiliation{Kobayashi-Maskawa Institute, Nagoya University, Nagoya 464-8602}
\affiliation{Nara Women's University, Nara 630-8506}
\affiliation{National Central University, Chung-li 32054}
\affiliation{National United University, Miao Li 36003}
\affiliation{Department of Physics, National Taiwan University, Taipei 10617}
\affiliation{H. Niewodniczanski Institute of Nuclear Physics, Krakow 31-342}
\affiliation{Nippon Dental University, Niigata 951-8580}
\affiliation{Niigata University, Niigata 950-2181}
\affiliation{Novosibirsk State University, Novosibirsk 630090}
\affiliation{Osaka City University, Osaka 558-8585}
\affiliation{Pacific Northwest National Laboratory, Richland, Washington 99352}
\affiliation{Panjab University, Chandigarh 160014}
\affiliation{University of Pittsburgh, Pittsburgh, Pennsylvania 15260}
\affiliation{Theoretical Research Division, Nishina Center, RIKEN, Saitama 351-0198}
\affiliation{University of Science and Technology of China, Hefei 230026}
\affiliation{Seoul National University, Seoul 151-742}
\affiliation{Showa Pharmaceutical University, Tokyo 194-8543}
\affiliation{Soongsil University, Seoul 156-743}
\affiliation{University of South Carolina, Columbia, South Carolina 29208}
\affiliation{Stefan Meyer Institute for Subatomic Physics, Vienna 1090}
\affiliation{Sungkyunkwan University, Suwon 440-746}
\affiliation{School of Physics, University of Sydney, New South Wales 2006}
\affiliation{Department of Physics, Faculty of Science, University of Tabuk, Tabuk 71451}
\affiliation{Tata Institute of Fundamental Research, Mumbai 400005}
\affiliation{Department of Physics, Technische Universit\"at M\"unchen, 85748 Garching}
\affiliation{Toho University, Funabashi 274-8510}
\affiliation{Department of Physics, Tohoku University, Sendai 980-8578}
\affiliation{Earthquake Research Institute, University of Tokyo, Tokyo 113-0032}
\affiliation{Department of Physics, University of Tokyo, Tokyo 113-0033}
\affiliation{Tokyo Institute of Technology, Tokyo 152-8550}
\affiliation{Tokyo Metropolitan University, Tokyo 192-0397}
\affiliation{University of Torino, 10124 Torino}
\affiliation{Virginia Polytechnic Institute and State University, Blacksburg, Virginia 24061}
\affiliation{Wayne State University, Detroit, Michigan 48202}
\affiliation{Yamagata University, Yamagata 990-8560}
\affiliation{Yonsei University, Seoul 120-749}
  \author{S.~Jia}\affiliation{Beihang University, Beijing 100191} 
  \author{C.~P.~Shen}\affiliation{Beihang University, Beijing 100191} 
  \author{C.~Z.~Yuan}\affiliation{Institute of High Energy Physics, Chinese Academy of Sciences, Beijing 100049} 
  \author{I.~Adachi}\affiliation{High Energy Accelerator Research Organization (KEK), Tsukuba 305-0801}\affiliation{SOKENDAI (The Graduate University for Advanced Studies), Hayama 240-0193} 
  \author{J.~K.~Ahn}\affiliation{Korea University, Seoul 136-713} 
  \author{H.~Aihara}\affiliation{Department of Physics, University of Tokyo, Tokyo 113-0033} 
  \author{S.~Al~Said}\affiliation{Department of Physics, Faculty of Science, University of Tabuk, Tabuk 71451}\affiliation{Department of Physics, Faculty of Science, King Abdulaziz University, Jeddah 21589} 
  \author{D.~M.~Asner}\affiliation{Pacific Northwest National Laboratory, Richland, Washington 99352} 
  \author{H.~Atmacan}\affiliation{University of South Carolina, Columbia, South Carolina 29208} 
  \author{T.~Aushev}\affiliation{Moscow Institute of Physics and Technology, Moscow Region 141700} 
% ...
\collaboration{The Belle Collaboration}

\begin{abstract}

We search for the $J^{PC}=0^{--}$  and $1^{+-}$ light
tetraquark states with masses up to 2.46~GeV/$c^2$
in $\Upsilon(1S)$ and $\Upsilon(2S)$ decays with data samples of
$(102\pm 2)$ million and $(158\pm 4)$ million events,
respectively, collected with the Belle detector. No significant
signals are observed in any of the studied production modes, and
90\% credibility level (C.L.) upper limits on their branching
fractions in $\Upsilon(1S)$ and $\Upsilon(2S)$ decays are
obtained.
The inclusive branching fractions of the $\Upsilon(1S)$ and
$\Upsilon(2S)$ decays into final states with $f_1(1285)$ are
measured to be $\BR(\Upsilon(1S)\to f_1(1285)+anything)=(46\pm28({\rm stat.})\pm13({\rm syst.}))\times 10^{-4}$ and
$\BR(\Upsilon(2S)\to f_1(1285)+anything)=(22\pm15({\rm stat.})\pm6.3({\rm syst.}))\times 10^{-4}$.
The measured $\chi_{b2} \to J/\psi + anything$ branching fraction is measured to be $(1.50\pm0.34({\rm stat.})\pm0.22({\rm syst.}))\times 10^{-3}$,
and 90\% C.L. upper limits for the $\chi_{b0,b1} \to J/\psi + anything$ branching fractions are found to be $2.3\times 10^{-3}$
and $1.1\times 10^{-3}$, respectively.
For $\BR(\chi_{b1} \to \omega + anything)$, the branching fraction is measured to be $(4.9\pm1.3({\rm stat.})\pm0.6({\rm syst.}))\times 10^{-2}$. 
All results reported here are the first measurements for these modes.
\end{abstract}


\maketitle

\section{\boldmath Introduction}
In the past decade, many experiments, both at lepton and
hadron colliders, have reported evidence for a large number of
particles having properties that can not be readily explained within the framework of the
expected heavy quarkonium states~\cite{A30.1530002, Eidel}. Among
them, the $X(3872)$~\cite{91.262001}, the
$Z_c(3900)$~\cite{110.252001,110.252002}, the
$X(3940)$~\cite{98.082001}, the
$Y(4260)$~\cite{95.142001,96.162003}, the
$Z(4430)$~\cite{112.222002}, the $Z_{b}(10610)$ and the
$Z_{b}(10650)$~\cite{108.122001}, are generally interpreted as possible tetraquark candidates with exotic
properties.

In the low-mass region, the Dalitz analysis of the decay
$D^{0}\to \pi^{+}\pi^{-}\pi^{0}$~\cite{99.251801} indicates the
existence of a state decaying into a $\rho\pi$ final state with exotic
quantum numbers $J^{PC}=0^{--}$~\cite{1257.242} at a mass of
$\approx$ 1865~MeV/$c^2$, which cannot be composed of a
quark-antiquark pair in the conventional quark
model~\cite{C38.090001,454.1}. If such a resonance exists, it
might be a hybrid or a tetraquark state~\cite{D92.114018}.

The authors of Ref.~\cite{1610.02081} calculated the masses of
such exotic four-quark states with $J^{PC}=0^{--}$ and $1^{+-}$ in
Laplace sum rules (LSR) and finite-energy sum rules (FESR) using
tetraquark-like currents. In the scalar channel, both LSR and FESR
gave consistent mass predictions of a tetraquark state with a mass
of $(1.66\pm 0.14)$~GeV/$c^2$. This numerical result favors the
tetraquark interpretation of the possible $\rho\pi$ dominance in
the $D^0$ decays. In the vector channel, the authors also
conservatively estimated the mass of a tetraquark state to be in
the mass region $1.18-1.43$~GeV/$c^2$. Although the masses have
been calculated, the width and couplings to any final
states were not predicted.

Very recently, the Belle Collaboration reported the search for the $J^{PC}=0^{--}$
glueball ($G_{0^{--}}$) in the production modes
$\Upsilon(1S,2S)\to \chi_{c1}+G_{0^{--}}$, $\Upsilon(1S,2S)\to
f_1(1285)+G_{0^{--}}$, $\chi_{b1}\to J/\psi+G_{0^{--}}$, and
$\chi_{b1}\to \omega+G_{0^{--}}$ with data samples of $(102\pm 2)$
million $\Upsilon(1S)$ and $(158\pm 4)$ million $\Upsilon(2S)$
events~\cite{oddball}. The masses of the putative glueballs
were fixed at 2.800, 3.810, and 4.330~GeV/$c^2$, as predicted from
quantum chromodynamics (QCD) sum rules~\cite{5} and distinct
bottom-up holographic models of QCD~\cite{1507}. Considering the
kinematical constraints and the conservation of the quantum numbers
$J^{PC}$, the production modes for glueball searches are also
suitable for searches for the aforementioned light tetraquark
states with $J^{PC}=0^{--}$ and $1^{+-}$, denoted collectively  as $X_{\rm
tetra}$.


In this paper, we utilize the low-mass recoil spectra of the
$\chi_{c1}$, $f_1(1285)$, $\jpsi$, and $\omega$ in bottomonium decays
to search for  $X_{\rm tetra}$ signals in the modes
$\Upsilon(1S,2S)\to \chi_{c1}+ X_{\rm tetra}$, $\Upsilon(1S,2S)\to
f_1(1285)+X_{\rm tetra}$, $\chi_{b1} \to J/\psi+X_{\rm tetra}$, and
$\chi_{b1}\to \omega+X_{\rm tetra}$~\cite{oddball}.
Since the $X_{\rm tetra}$ properties are unknown, we report our investigation for
different assumed values for the $X_{\rm tetra}$ mass and width.

As byproducts of the $X_{\rm tetra}$ search, we measure the inclusive $f_1(1285)$
production in $\Upsilon(1S,2S)$, $J/\psi$ production in $\chi_{bJ} (J = 0, 1, 2)$, and $\omega$ production in $\chi_{b1}$ decays.

\section{\boldmath The Data Sample and Belle Detector}

This analysis utilizes the Belle $\Upsilon(1S)$ and $\Upsilon(2S)$ data
samples with a total luminosity of $5.74$ and
$24.91~{\rm fb}^{-1}$, respectively,
corresponding to $(102\pm2)\times 10^6$ $\Upsilon(1S)$ and $(158\pm4)\times 10^6$
$\Upsilon(2S)$ events~\cite{number}.
An $89.45~{\rm fb}^{-1}$ data sample collected at $\sqrt{s} =
10.52~\mathrm{GeV}$ is used to estimate the possible irreducible
contributions from continuum ($e^+ e^- \to q\bar{q}$, where $q \in \{u,\,d,\,s,\,c\}$). Here, $\sqrt{s}$ is the center-of-mass
(C.M.) energy of the colliding $e^+e^-$ system. The data were
collected with the Belle
detector~\cite{Abashian2002117,PTEP201204D001} operated at the
KEKB asymmetric-energy
$e^{+}e^{-}$~collider~\cite{Kurokawa20031,PTEP201303A001}. Large
Monte Carlo (MC) samples of all of the investigated tetraquark modes
are generated with {\sc evtgen}~\cite{Lange2001152}
and simulated with a GEANT3-based~\cite{geant3} model for the detector response
to determine the
signal line-shapes and efficiencies.
The angular distribution for the decay
$\Upsilon(2S) \to \gamma \chi_{bJ}$ is simulated assuming a pure
$\rm{E1}$ transition ($dN/d\cos\theta_{\gamma} \propto 1+ \alpha
\cos^2\theta_{\gamma}$ with $\alpha=1$, $-\frac{1}{3}$, $\frac{1}{13}$
for $J=0$, 1, 2, respectively~\cite{mcang}, where $\theta_\gamma$ is the
polar angle of the $\Upsilon(2S)$ radiative photon in the $e^+e^-$
C.M. frame);  a phase space model in {\sc evtgen} is used for the $\chi_{bJ}$
decays. We use the phase space model for other decays
as well. Note that the $X_{\rm tetra}$ inclusive decays are modelled using
{\sc pythia}~\cite{JHEP2006.026}. Inclusive $\Upsilon(1S)$ and
$\Upsilon(2S)$ MC samples, produced using {\sc
pythia} with four times the total numbers of $\Upsilon(1S,2S)$ events
of the data, are used to identify possible backgrounds showing peak distributions
from $\Upsilon(1S)$ and $\Upsilon(2S)$ decays.

The Belle detector is a large solid-angle magnetic spectrometer
that consists of a silicon vertex detector, a 50-layer central
drift chamber, an array of aerogel threshold Cherenkov
counters, a barrel-like arrangement of time-of-flight
scintillation counters, and an electromagnetic calorimeter
comprised of CsI(Tl) crystals located inside a
superconducting solenoid coil that provides a $1.5~\hbox{T}$
magnetic field.
An iron flux-return yoke instrumented with resistive plate chambers
located outside the coil
is used to detect $K^{0}_{L}$ mesons and to identify
muons. A detailed description of the Belle detector can be found
in Refs.~\cite{Abashian2002117, PTEP201204D001}.

\section{\boldmath MEASUREMENTS OF $\Upsilon(1S,2S\bf) \to f_1(1285)+ anything$}

\begin{figure*}[htbp]
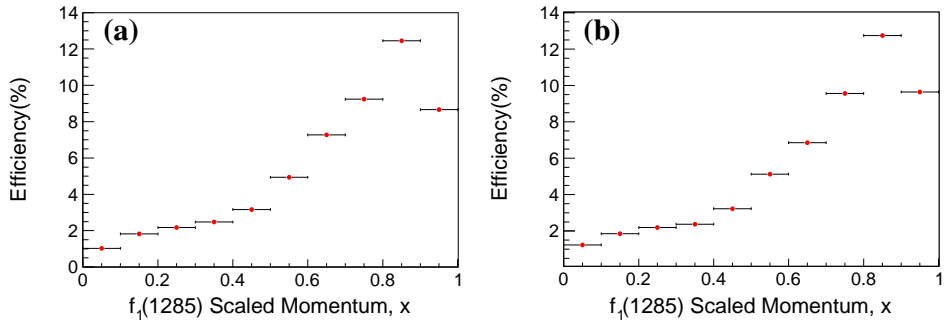

\includegraphics[height=6cm,angle=-90]{fig1a.epsi}
\hspace{0.2cm}
\includegraphics[height=6cm,angle=-90]{fig1b.epsi} 
\caption{
MC efficiencies for reconstructed $f_1(1285)$ mesons in (a) $\Upsilon(1S)$ and (b)
$\Upsilon(2S)$ decays as a function of the scaled momentum $x$.} \label{f1f2eff}
\end{figure*}

Candidate $f_1(1285)$ states are reconstructed via $\eta\pi^+\pi^-$, $\eta\to \gamma \gamma$.
Considering the differences in the MC-determined reconstruction efficiencies for different $f_1(1285)$ momenta, we
partition the data samples according to the scaled momentum $x=2\sqrt{s} \times p^{\ast}_{f_1(1285)}/(s-m_{f_1(1285)}^2)$,
where $p^{\ast}_{f_1(1285)}$ is the momentum of the $f_1(1285)$ candidate in the C.M. system, and $m_{f_1(1285)}$ is the
$f_1(1285)$ nominal mass~\cite{C38.090001}. The normalizing expression $(s-m_{f_1(1285)}^2)/(2\sqrt{s})$
represents  the maximum value of $p^{\ast}_{f_1(1285)}$ for the case where the $f_1(1285)$ candidate recoils against a massless particle.
The use of $x$
removes the beam-energy dependence in comparing the continuum data
to those taken at the $\Upsilon(1S,2S)$ resonances.
The event selections are identical to those used in Ref.~\cite{oddball}. Figure~\ref{f1f2eff}
shows the reconstruction efficiencies as a function of $x$ for $f_1(1285)$ candidates from $\Upsilon(1S,2S)$ decays
in each $x$ interval. Here, the efficiencies are
estimated using a MC signal sample generated on the basis of the relative weights of the differential
branching fractions (discussed below) in the different $x$ bins.

The invariant mass distributions for the $f_1(1285)$ candidates in $\Upsilon(1S,2S)$ data for the entire $x$ region and
for subranges in $x$ are shown in
Figs.~\ref{f1datasamples} and ~\ref{f2datasamples}.
We observe clear $f_1(1285)$ signals in high-$x$ bins and $\eta(1405)$ signals in the subregion $0.6 < x < 1.0$.
In the figures, the cross-hatched histograms are from the normalized continuum contributions.
See Ref.~\cite{oddball} for
the definition of the normalization method of the continuum contribution.
For $\Upsilon(2S)\to f_1(1285)+anything$, a further background arises from the intermediate transition $\Upsilon(2S)
\to \pi^+ \pi^- \Upsilon(1S)$ or $\pi^0 \pi^0 \Upsilon(1S)$ with $\Upsilon(1S)$ decaying to $f_1(1285)$. This
contamination is removed by requiring the $\pi \pi$ recoil mass to be outside the $[9.45, 9.47]$~GeV/$c^2$ range for all $\pi\pi$ combinations~\cite{oddball}.

\begin{figure}[htpb]
\includegraphics[height=6cm, angle=-90]{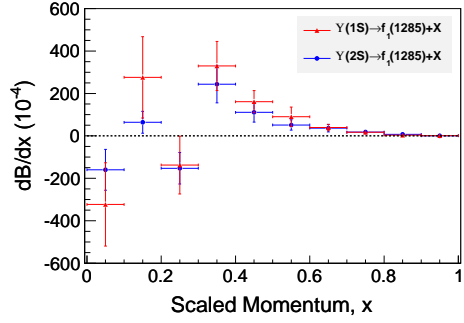}
\caption{(Color online) Differential branching fractions for $\Upsilon(1S)$ and
$\Upsilon(2S)$ inclusive decays into $f_1(1285)$ as a function of the scaled momentum $x$ defined in the text.
The error bar of each point is the sum of the statistical and systematic errors.}\label{differential}
\end{figure}

A binned extended simultaneous likelihood fit is applied to the $x$-dependent $\eta \pi^+ \pi^-$ invariant mass spectra to extract the $f_1(1285)$
signal yields in the $\Upsilon(1S,2S)$ and continuum data samples. Due to the dependence on momentum, the $f_1(1285)$ and $\eta(1405)$ signal shapes in each
$x$ bin are described by Voigtian functions (a Breit-Wigner distribution convolved with a Gaussian function) that are obtained from the MC simulations directly; a third-order
Chebyshev polynomial background shape is used for the $\Upsilon(1S,2S)$ decay backgrounds in addition to the normalized continuum contributions.
The fit results are shown in Figs.~\ref{f1datasamples} and \ref{f2datasamples} for the $\Upsilon(1S)$ and $\Upsilon(2S)$ decays, respectively. The fitted $f_1(1285)$ signal yields ($N_{\rm fit}$) in each $x$ bin from $\Upsilon(1S)$ and $\Upsilon(2S)$ decays are tabulated in Table~\ref{Table-Nfit}, together with the reconstruction efficiencies from MC signal simulations ($\varepsilon$), the total systematic uncertainties ($\sigma_{\rm syst}$) discussed below (which are the sum of the common systematic errors, fit uncertainties and
continuum-scale-factor uncertainties), and the corresponding branching fractions ($\BR$). The total numbers of $f_1(1285)$ events, \textit{i.e.}, the sums of the signal yields in all of the $x$ bins, the sums of the $x$-dependent efficiencies weighted by the signal fraction in that $x$ bin, and the measured branching fractions are listed in the bottom row of Table~\ref{Table-Nfit}.
The branching fractions for $\Upsilon(1S,2S)\to f_1(1285) + anything$ are measured to be:
$$
\BR(\Upsilon(1S)\to f_1(1285)+
anything)=(46\pm28({\rm stat.})\pm13({\rm syst.}))\times 10^{-4},
$$
$$
\BR(\Upsilon(2S)\to f_1(1285)+
anything)=(22\pm15({\rm stat.})\pm6.3({\rm syst.}))\times 10^{-4}.
$$
The differential branching fractions of $\Upsilon(1S,2S)$ decays to $f_1(1285)$ are shown in Fig.~\ref{differential}.

\begin{figure*}[htbp]
\includegraphics[height=4.1cm,angle=-90]{fig2a.epsi}
\hspace{0.15cm}
\includegraphics[height=4.1cm,angle=-90]{fig2b.epsi}
\hspace{0.15cm}
\includegraphics[height=4.1cm,angle=-90]{fig2c.epsi}
\vspace{0.2cm}

\includegraphics[height=3.8cm,angle=-90]{fig2d.epsi}
\hspace{0.15cm}
\includegraphics[height=3.8cm,angle=-90]{fig2e.epsi}
\hspace{0.15cm}
\includegraphics[height=3.8cm,angle=-90]{fig2f.epsi}
\hspace{0.15cm}
\includegraphics[height=3.8cm,angle=-90]{fig2g.epsi}
\vspace{0.2cm}

\includegraphics[height=3.8cm,angle=-90]{fig2h.epsi}
\hspace{0.15cm}
\includegraphics[height=3.8cm,angle=-90]{fig2i.epsi}
\hspace{0.15cm}
\includegraphics[height=3.8cm,angle=-90]{fig2j.epsi}
\hspace{0.15cm}
\includegraphics[height=3.8cm,angle=-90]{fig2k.epsi}
\caption{(Color online) Invariant mass distributions of the $f_1(1285)$
candidates in (a) the entire $x$ region and (b-k) for $x$ bins of size
$0.1$. The dots with error bars are the $\Upsilon(1S)$
data. The red solid lines are the best fits, and the blue dotted lines represent the total backgrounds.
The cross-hatched green histograms are from the normalized
continuum contributions.}\label{f1datasamples}
\end{figure*}

\begin{figure*}[htbp]
\includegraphics[height=4.0cm,angle=-90]{fig3a.epsi}
\hspace{0.15cm}
\includegraphics[height=4.0cm,angle=-90]{fig3b.epsi}
\hspace{0.15cm}
\includegraphics[height=4.0cm,angle=-90]{fig3c.epsi}
\vspace{0.2cm}

\includegraphics[height=3.8cm,angle=-90]{fig3d.epsi}
\hspace{0.15cm}
\includegraphics[height=3.8cm,angle=-90]{fig3e.epsi}
\hspace{0.15cm}
\includegraphics[height=3.8cm,angle=-90]{fig3f.epsi}
\hspace{0.15cm}
\includegraphics[height=3.8cm,angle=-90]{fig3g.epsi}
\vspace{0.2cm}

\includegraphics[height=3.8cm,angle=-90]{fig3h.epsi}
\hspace{0.15cm}
\includegraphics[height=3.8cm,angle=-90]{fig3i.epsi}
\hspace{0.15cm}
\includegraphics[height=3.8cm,angle=-90]{fig3j.epsi}
\hspace{0.15cm}
\includegraphics[height=3.8cm,angle=-90]{fig3k.epsi}
\caption{(Color online) Invariant mass distributions of the $f_1(1285)$
candidates in (a) the entire $x$ region and (b-k) for $x$ bins of size
$0.1$. The dots with error bars are the $\Upsilon(2S)$
data. The red solid lines are the best fits, and the blue dotted lines represent the total backgrounds.
The green cross-hatched histograms are from the normalized
continuum contributions.}\label{f2datasamples}
\end{figure*}

\begin{table*}[t]
\begin{threeparttable}
\caption{\label{Table-Nfit} Summary of the branching fraction
measurements of $\Upsilon(1S,2S)$ inclusive decays into
$f_1(1285)$, where $N_{\rm fit}$ is the number of fitted signal
events, $\varepsilon$ is the reconstruction efficiency,
$\sigma_{\rm syst}$ is the relative total systematic uncertainty, and $\BR$ is the measured branching fraction.}
\scriptsize
  \begin{tabular}{cr@{$\pm$}lcccr@{$\pm$}lccc}
  \hline\hline
    \multicolumn{6}{c}{$\Upsilon(1S)\to f_1(1285)+{\rm anything}$} & \multicolumn{5}{c}{$\Upsilon(2S)\to f_1(1285)+{\rm anything}$} \\[-2pt]
    \specialrule{0em}{1.5pt}{1.5pt}
    \vspace{0.1cm}
   $x$ & \multicolumn{2}{c}{$N_{\rm fit}$} & $\varepsilon (\%)$ & $\sigma_{\rm syst} (\%)$ & $\mathcal{B}(10^{-4})$& \multicolumn{2}{c}{$N_{\rm fit}$} & $\varepsilon (\%)$ & $\sigma_{\rm syst} (\%)$ & $\mathcal{B}(10^{-4})$ \\
  \hline
  \specialrule{0em}{1pt}{1pt}
  \vspace{0.1cm}
$(0.0,0.1)$  &$-480$  & $239$ & $1.03$ &$24.5$ &$-32\pm16\pm8.0$ & $-442$  & $253$ & $1.23$ &$29.8$ &  $-16\pm9.2\pm4.8$ \\
\vspace{0.1cm}
$(0.1,0.2)$  &$727$ & $497$ & $1.82$ &$25.5$  &$28\pm19\pm7.1$& $265$  & $192$ & $1.85$  &$26.9$ & $6.4\pm4.7\pm1.8$ \\
\vspace{0.1cm}
$(0.2,0.3)$  &$-432$  & $339$ & $2.17$ &$24.6$  &$-14\pm11\pm3.4$ & $-749$  & $333$ & $2.19$  &$26.0$ &$-15\pm6.8\pm4.0$ \\
\vspace{0.1cm}
$(0.3,0.4)$  & $1181$  & $240$ & $2.48$&$28.9$ &$33\pm6.7\pm9.6$ & $1296$  & $348$ & $2.37$  &$25.3$ &$24\pm6.6\pm6.2$ \\
\vspace{0.1cm}
$(0.4,0.5)$  & $736$  & $165$ & $3.16$ &$24.2$ &$16\pm3.6\pm3.9$ & $801$  & $247$ & $3.22$  &$26.7$& $11\pm3.5\pm3.0$ \\
\vspace{0.1cm}
$(0.5,0.6)$  & $645$  & $126$ & $4.94$&$36.4$ &$9.0\pm1.8\pm3.3$ & $590$  & $189$ & $5.12$  &$34.9$ & $5.1\pm1.7\pm1.8$ \\
\vspace{0.1cm}
$(0.6,0.7)$  & $412$  & $88$ & $7.27$&$31.3$ &$3.9\pm0.9\pm1.3$ & $563$  & $143$ & $6.86$  &$32.6$ & $3.7\pm1.0\pm1.2$ \\
\vspace{0.1cm}
$(0.7,0.8)$  & $229$  & $65$ & $9.24$ &$42.8$ &$1.7\pm0.5\pm0.8$ & $382$  & $70$ & $9.56$ &$35.6$ & $1.8\pm0.4\pm0.7$ \\
\vspace{0.1cm}
$(0.8,0.9)$  & $66$  & $38$ & $12.46$ &$48.0$ &$0.4\pm0.3\pm0.2$ & $205$  & $84$ & $12.75$ &$36.3$ & $0.7\pm0.3\pm0.3$ \\
\vspace{0.1cm}
$(0.9,1.0)$  & $16$  & $11$ & $8.66$ &$55.0$ &$0.1\pm0.1\pm0.1$ & $15$  & $11$ & $9.65$  &$48.9$ & $0.1\pm0.1\pm0.1$ \\
All $x$  &$3100$ &$950$ & $4.68$  & $28.7$ &$46\pm28\pm13$ &$2926$ &$712$ &$5.93$  & $28.4$ & $22\pm15\pm6.3$ \\
  \hline\hline
  \end{tabular}
  \end{threeparttable}
\end{table*}

\section{\boldmath MEASUREMENTS OF $\chi_{bJ} \to J/\psi +  {\rm anything}$}
\begin{figure}[htpb]
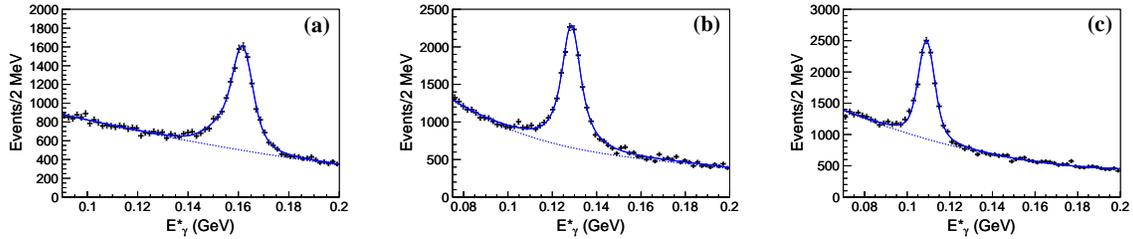

\includegraphics[height=4.5cm, angle=-90]{fig5a.epsi}
\hspace{0.5cm}
\includegraphics[height=4.5cm, angle=-90]{fig5b.epsi}
\hspace{0.5cm}
\includegraphics[height=4.5cm, angle=-90]{fig5c.epsi}
\caption{The spectra of the $\Upsilon(2S)$ radiative photon energy in the $\EE$ C.M. frame from MC simulated
$\Upsilon(2S)\to \gamma \chi_{bJ}$, $\chi_{bJ} \to J/\psi + anything$ signal samples for (a) $\chi_{b0}$, (b)
$\chi_{b1}$ and (c) $\chi_{b2}$, respectively.}\label{2Smcgam}
\end{figure}

The $\chi_{bJ}$ is identified through the decay $\Upsilon(2S)\to \gamma \chi_{bJ}$.
The same mass regions of the $J/\psi$ signal and sidebands are used as in Ref.~\cite{oddball}, \textit{i.e.},
we define the $\jpsi$ signal region to be the window
$|M_{\ell^+\ell^-}-m_{\jpsi}|<0.03$~GeV/$c^2$ ($\sim 2.5\sigma$), where $m_{\jpsi}$ is the $\jpsi$ nominal mass~\cite{C38.090001}, while the $\jpsi$ sideband is $2.97~\hbox{GeV}/c^2<M_{\ell^+\ell^-}<3.03$~GeV/$c^2$ or $3.17~\hbox{GeV}/c^2<M_{\ell^+\ell^-}<3.23$~GeV/$c^2$, which is twice as wide as
the signal region. After requiring the lepton-pair mass to be within the $\jpsi$ signal region, Figs.~\ref{2Smcgam} (a--c) show the distributions of the $\Upsilon(2S)$
radiative photon energy in the $\EE$ C.M. frame from MC simulated $\Upsilon(2S)\to \gamma \chi_{bJ}$, $\chi_{bJ} \to J/\psi + anything$ decays,
where each $\chi_{bJ}$ signal shape is described by the convolution of a BW function
with a Novosibirsk~\cite{Nov} function. Based on the fitted results, the efficiencies are $(23.87\pm0.42)\%$, $(32.21\pm0.53)\%$, and $(22.96\pm0.39)\%$ for $\chi_{b0}$, $\chi_{b1}$ and $\chi_{b2}$, respectively.

\begin{figure*}[htbp]
\includegraphics[height=7cm,angle=-90]{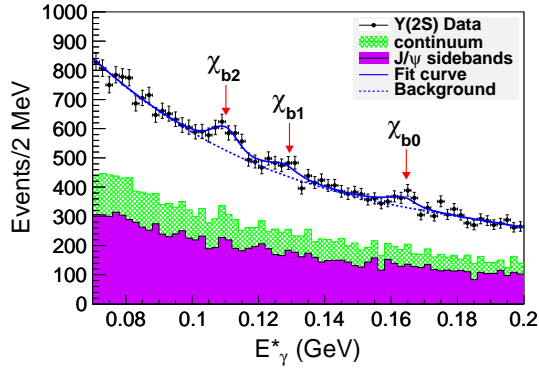}
\caption{(Color online)
The spectra of the $\Upsilon(2S)$ radiative photon energy in the $\EE$ C.M. frame in $\Upsilon(2S)$ data. The dots with error bars are the $\Upsilon(2S)$
data. The blue solid line is the best fit, and the blue dotted line represents the backgrounds. The magenta  shaded histogram is from the normalized $\jpsi$
sideband and the green  cross-hatched histogram is from the normalized continuum contributions described in the text.}
\label{2Sdatagam}
\end{figure*}

As shown in Fig.~\ref{2Sdatagam} of the spectrum of the $\Upsilon(2S)$ radiative photon energy in the C.M. frame,
a clear $\chi_{b2}$ signal may be observed. After all selection requirements, no backgrounds showing peak distributions are found
in the distribution estimated from $\jpsi$ mass sideband data, nor in the
continuum production in the $\chi_{bJ}$ signal regions, in agreement with the expectation from the $\Upsilon(2S)$ generic MC samples.
An unbinned extended maximum-likelihood fit to the spectrum is performed to extract the signal and background yields in the $\Upsilon(2S)$ data samples.
In the fit, the probability density function (PDF) of each $\chi_{bJ}$ signal is a
BW function convolved with a Novosibirsk function with all the parameters free; for the background PDF, a third-order Chebyshev polynomial function is adopted. The fit yields $243\pm101$, $269\pm120$, and $462\pm105$ events for the $\chi_{b0}$, $\chi_{b1}$, and $\chi_{b2}$ signals, respectively, in the $\Upsilon(2S)$ data sample.
The statistical significances of the $\chi_{b0}$, $\chi_{b1}$ and $\chi_{b2}$ signals are estimated to be $1.5\sigma$, $1.1\sigma$ and $3.5\sigma$, from the differences of the logarithmic likelihoods, $-2\ln(\mathcal{L}_0/\mathcal{L}_{\rm max})$, where $\mathcal{L}_{\rm 0}$ and $\mathcal{L}_{\rm max}$ are the likelihoods of the fits without and with a signal component, respectively (taking the number of degrees of freedom in each fit into account).
For $\chi_{b2} \to J/\psi + anything$, the branching fraction is measured for the first time using
$$
\BR(\chi_{b2} \to J/\psi + anything)=
\frac{N_{\chi_{b2}}}{N_{\Upsilon(2S)}\times
\varepsilon_{\chi_{b2}}\times \BR(\Upsilon(2S)\to \gamma\chi_{b2})\times \BR(J/\psi \to \ell^{+} \ell^{-})},
$$
where $N_{\chi_{b2}}$ is the number of fitted $\chi_{b2}$ signal events and $\varepsilon_{\chi_{b2}}$ is the signal detection
efficiency given above.
We measure a value of $(1.50\pm0.34({\rm stat.})\pm0.22({\rm syst.}))\times 10^{-3}$.
The systematic uncertainties are discussed below.
The $\chi_{b0,b1}$ branching fractions are computed in a similar way. Since the $\chi_{b0,b1}$ signal significances are less than $3\sigma$,
we compute 90\% credibility level (C.L.) upper limits $x^{\rm UL}$ on the $\chi_{b0,b1}$ signal yields and the branching fractions.
For this purpose, we solve the equation $\int^{x^{\rm UL}}_0 \mathcal{L}(x)dx / \int^{+\infty}_0\mathcal{L}(x)dx = 0.9$, where $x$ is the
assumed signal yield or branching fraction, and  $\mathcal{L}(x)$ is the corresponding likelihood of the data.
To take into account the systematic uncertainties discussed below, the likelihood is convolved with a Gaussian function whose width equals the
total systematic uncertainty. The upper limits for the yields of $\chi_{b0}$ and $\chi_{b1}$ are 380 and 432 respectively, and the corresponding
upper limits on the branching fractions are $\BR^{\rm UL}(\chi_{b0} \to J/\psi+ anything) = 2.3\times 10^{-3}$ and $\BR^{UL}(\chi_{b1} \to J/\psi+anything)
= 1.1\times10^{-3}$ at 90\% C.L.

\section{\boldmath MEASUREMENTS OF $\chi_{b1} \to \omega +  {\rm anything}$}
\begin{figure*}[htbp]
\includegraphics[height=7cm,angle=-90]{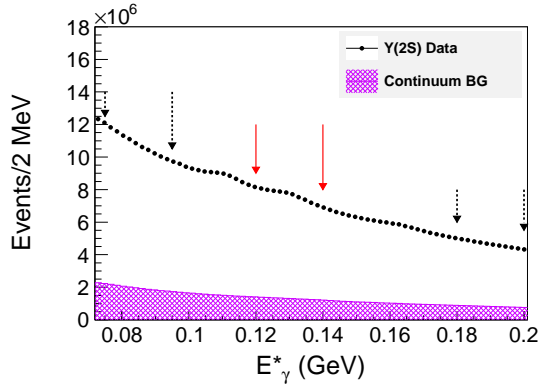}
\caption{(Color online)
The spectra of the $\Upsilon(2S)$ radiative photon energy in the $\EE$ C.M. frame, where the dots with imperceptible error bars are the $\Upsilon(2S)$
data and the magenta  cross-hatched histogram is from the normalized continuum contributions.
The red solid arrows indicate the selected $\chi_{b1}$ signal region, and the black dashed
arrows show the two ranges of the $\chi_{b1}$ sideband.}
\label{omechib1}
\end{figure*}

Candidate $\omega$ mesons are reconstructed via $\pp\pi^0$.
We perform a mass-constrained kinematic fit to the selected $\pi^0$ candidate
and require $\chi^{2}<10$. To remove the backgrounds with
$K^{0}_{S}$, the $\pi^+\pi^-$ invariant mass is required
to be outside the [0.475, 0.515]~GeV/$c^2$ range. After requiring the $\pp\pi^0$ invariant mass to be within the
$\omega$ signal region of $0.755~\hbox{GeV}/c^2<M(\pi^+\pi^-\pi^0)<0.805$~GeV/$c^2$, Fig.~\ref{omechib1} shows the
distributions of the energy of the $\Upsilon(2S)$ radiative photon in the C.M. frame,
where the dots represent the $\Upsilon(2S)$ data and the cross-hatched histogram is from the normalized
continuum contributions.
We define the $\chi_{b1}$ signal region as $0.12~\hbox{GeV}<E^{*}_{\gamma}<0.14~\hbox{GeV}$ and its sideband
as $0.075~\hbox{GeV}<E^{*}_{\gamma}<0.095~\hbox{GeV}$ or $0.18~\hbox{GeV}<E^{*}_{\gamma}<0.20~\hbox{GeV}$, which is twice as wide as
the signal region. From the histogram, no $\chi_{b1}$ signal is present in the continuum contributions.

\begin{figure*}[htbp]
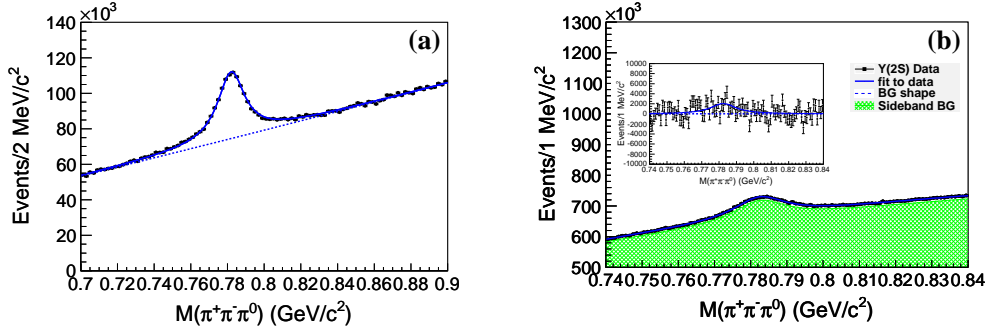

\includegraphics[height=6cm,angle=-90]{fig8a.epsi}
\hspace{0.8cm}
\includegraphics[height=6cm,angle=-90]{fig8b.epsi}
\caption{(Color online) The $\pp\pi^0$ invariant mass spectra
from (a) MC simulated $\chi_{b1} \to \omega + anything$ signal sample and
(b) $\Upsilon(2S)$ data. The dots represent the data. The cross-hatched histogram in (b) represents the normalized $\chi_{b1}$
sideband; the inset shows the fitted background-subtracted distribution.
The blue solid lines are the best fits, and the blue dotted lines represent the backgrounds.}
\label{omeM}
\end{figure*}

After the application of the above requirements, the $\pp\pi^0$ invariant mass distribution from MC simulated $\chi_{b1} \to \omega + anything$
signal sample is shown in Fig.~\ref{omeM}a.
In the fit to this distribution, a Voigtian function is used for the $\omega$ signal shape and
a second-order Chebyshev polynomial function is used for the
background shape. Based on the fitted result, the efficiency is $(10.9\pm0.1)\%$.
Figure~\ref{omeM}b shows the distributions of the $\pp\pi^0$ invariant mass from the $\Upsilon(2S)$ data (the dots with error bars)
and the normalized $\chi_{b1}$ sideband events (the cross-hatched histogram).
From the plot, the observed $\omega$ signals in the normalized $\chi_{b1}$ sideband account for most of the
events in the $\chi_{b1}$ signal region.

A simultaneous binned extended  maximum likelihood fit is applied to the $\pp\pi^0$ invariant mass spectra
to extract the $\omega$ signal yields in the $\chi_{b1}$ signal region and its sideband. The $\omega$ signal shape is
described by a Voigtian function with the values of the parameters fixed to those from the
fit to MC-simulated signals; a second-order Chebyshev polynomial background shape is used for the $\chi_{b1}$ decay backgrounds
in addition to the normalized $\chi_{b1}$ sideband. The fitted $\omega$ signal yield is $51054\pm12943$
and the estimated statistical significance is $4.1\sigma$.
Hence, the branching fraction for $\chi_{b1} \to \omega + anything$ is measured for the first time to be
$$
\BR(\chi_{b1} \to \omega + anything)=(4.9\pm1.3({\rm stat.})\pm0.6({\rm syst.}))\times 10^{-2}.
$$


\section{\boldmath Search for $X_{\rm tetra}$ in $\Upsilon(1S)$, $\Upsilon(2S)$, and $\chi_{b1}$ decays}

We generate a large number of MC samples for $\Upsilon(1S,2S)\to
\chi_{c1}+X_{\rm tetra}$, $\Upsilon(1S,2S)\to f_1(1285)+X_{\rm
tetra}$, $\chi_{b1}\to J/\psi+X_{\rm tetra}$, and $\chi_{b1} \to
\omega+X_{\rm tetra}$ with $X_{\rm tetra}$ masses varying from
1.16 to 2.46~GeV/$c^2$ in steps of 0.10~GeV/$c^2$ and widths
varying from 0.0 to 0.3~GeV in steps of 0.1~GeV, using the same
decay modes as in Ref.~\cite{oddball}. After applying all the
event selections in Ref.~\cite{oddball}, all relevant
efficiencies are obtained; they are displayed graphically in
Fig.~\ref{eff}. Since the event selection requirements are
independent of the recoil part of the $\chi_{c1}$, $f_1(1285)$,
$\jpsi$, and $\omega$ in the studied channels, the detection
efficiencies are only related to the recoil masses. The
efficiencies versus $X_{\rm tetra}$ mass in the entire region from
1.16 to 3.0~GeV/$c^2$  are displayed graphically
in Fig.~\ref{eff} for the studied production modes.
The fitted curves show the second-order Chebyshev polynomials used to model these efficiencies.

\begin{figure*}[htbp]
\includegraphics[height=5cm,angle=-90]{fig9a.epsi}
\hspace{0.2cm}
\includegraphics[height=5cm,angle=-90]{fig9b.epsi}
\hspace{0.2cm}
\includegraphics[height=5cm,angle=-90]{fig9c.epsi}
\\\vspace{0.2cm}
\includegraphics[height=5cm,angle=-90]{fig9d.epsi}
\hspace{0.2cm}
\includegraphics[height=5cm,angle=-90]{fig9e.epsi}
\hspace{0.2cm}
\includegraphics[height=5cm,angle=-90]{fig9f.epsi}
\caption{(Color online)
Reconstruction efficiencies for (a) $\Upsilon(1S)\to \chi_{c1}+X_{\rm tetra}$,
(b) $\Upsilon(2S)\to \chi_{c1}+X_{\rm tetra}$, (c)
$\Upsilon(1S)\to f_1(1285)+X_{\rm tetra}$, (d) $\Upsilon(2S)\to
f_1(1285)+X_{\rm tetra}$, (e) $\chi_{b1}\to J/\psi+X_{\rm tetra}$
and (f) $\chi_{b1}\to \omega+X_{\rm tetra}$ as a function of the
assumed $X_{\rm tetra}$ masses, with $X_{\rm tetra}$ widths varying
from 0.0 to 0.3~GeV in steps of 0.1~GeV. The four solid lines in each panel, one for each $X_{\rm tetra}$ width,  are the
fits of a second-order Chebyshev polynomial to these data.}\label{eff}
\end{figure*}

In the channels analyzed below, $\Upsilon(1S,2S)\to
\chi_{c1}+X_{\rm tetra}$, $\Upsilon(1S,2S)\to f_1(1285)+X_{\rm
tetra}$, $\chi_{b1}\to J/\psi+X_{\rm tetra}$, and $\chi_{b1}\to
\omega+X_{\rm tetra}$, we search for the $X_{\rm tetra}$ signals
in the recoil mass spectra of the $\chi_{c1}$, $f_1(1285)$,
$\jpsi$, and $\omega$, respectively, with $X_{\rm tetra}$ widths
between  0.0 and 0.3~GeV in steps of 0.1~GeV. All recoil
mass spectra are taken from Ref.~\cite{oddball} with a focused view
of the low-mass region.

\begin{figure*}[htbp]
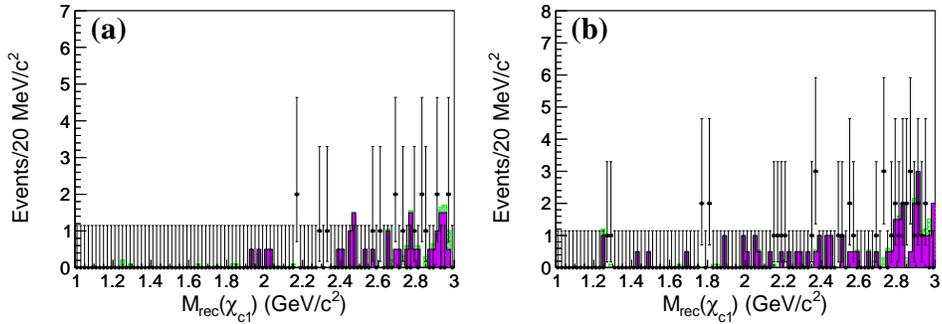

\includegraphics[height=6cm,angle=-90]{fig10a.epsi}
\hspace{0.2cm}
\includegraphics[height=6cm,angle=-90]{fig10b.epsi}
\caption{(Color online) The $\chi_{c1}$ recoil mass spectra in the
(a) $\Upsilon(1S)$ and (b) $\Upsilon(2S)$ data samples. The shaded
histograms are from the normalized $\chi_{c1}$ sideband
and the cross-hatched histograms show the normalized
continuum contributions~\cite{oddball}.} \label{Y12Sdata}
\end{figure*}

For $\Upsilon(1S,2S)\to \chi_{c1}+X_{\rm tetra}$, the $\chi_{c1}$
is reconstructed via its decay into $\gamma \jpsi$, $\jpsi\to
\LL$ ($\ell=e$ or $\mu$). Figure~\ref{Y12Sdata} shows the recoil
mass spectra of $\chi_{c1}$ candidates in the $\Upsilon(1S,2S)$
data, where the shaded histograms are from the normalized
$\chi_{c1}$ sideband and the cross-hatched histograms
show the normalized continuum contributions. See Ref.~\cite{oddball} for  the definition of
the $\chi_{c1}$  sideband and the normalization method of the
continuum contribution. There
are no evident signals for any of the $X_{\rm tetra}$ states at
any of the masses. In the entire region of study, the most significant signal is observed
at an $X_{\rm tetra}$ mass of 2.46 (2.26)~GeV/$c^2$ and width of 0.3 (0.0) GeV
with a statistical significance of 1.4$\sigma$ (0.6$\sigma$) in $\Upsilon(1S)$ $(\Upsilon(2S))$ data.
Since the number of selected signal
candidate events is small, we obtain the 90\%
C.L. upper limit of the signal yield ($N^{\rm UL}$) at each
$X_{\rm tetra}$ mass point by using the frequentist
approach~\cite{D57.3873} implemented in the POLE (Poissonian limit
estimator) program~\cite{D67.012002}, where each mass region is
selected to contain 95\% of the signal according to MC
simulations, the number of observed signal events is counted
directly, and the number of expected background events is estimated
from the sum of the normalized $\chi_{c1}$ sideband and
continuum contributions. The systematic uncertainties discussed
below are taken into account.

\begin{figure*}[htbp]
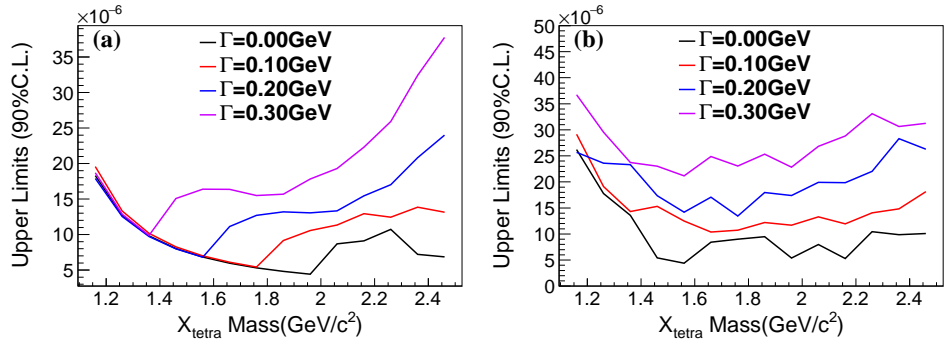

\includegraphics[height=6cm,angle=-90]{fig11a.epsi}
\hspace{0.3cm}
\includegraphics[height=6cm,angle=-90]{fig11b.epsi}
\caption{ (Color online) The upper limits on the branching
fractions for (a) $\Upsilon(1S)\to \chi_{c1}+X_{\rm tetra}$ and
(b) $\Upsilon(2S)\to \chi_{c1}+X_{\rm tetra}$ as a function of the
assumed $X_{\rm tetra}$ mass with widths fixed at 0.0, 0.1, 0.2,
and 0.3 GeV.}\label{final1}
\end{figure*}

The calculated upper limits on the numbers of signal events
($N^{\rm UL}$) and branching fraction ($\BR^{\rm UL}$) for each
$X_{\rm tetra}$ state with $X_{\rm tetra}$ masses from 1.16 to
2.46~GeV/$c^2$ and widths from 0.0 to 0.3~GeV in $\Upsilon(1S,2S)$
data are listed in Table~\ref{Summary}, together with the
reconstruction efficiencies ($\varepsilon$) and the systematic
uncertainties ($\sigma_{\rm syst}$). The results are displayed
graphically in Fig.~\ref{final1}.

\begin{figure*}[htbp]
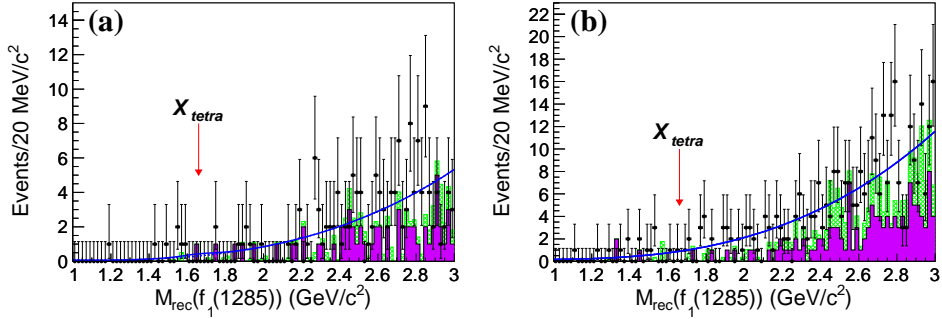

\includegraphics[height=6cm,angle=-90]{fig12a.epsi}
\hspace{0.2cm}
\includegraphics[height=6cm,angle=-90]{fig12b.epsi}
\caption{(Color online) The $f_1(1285)$ recoil mass spectra in the
(a) $\Upsilon(1S)$ and (b) $\Upsilon(2S)$ data samples. The
blue solid curves show the results of the fit described in the text,
including the $X_{\rm tetra}$ states with widths fixed at 0.10~GeV
and masses fixed at 1.66~GeV/$c^2$ indicated by the arrows. The nearly imperceptible blue
dashed curves show the fitted background. The magenta  shaded histograms
are from the normalized $f_1(1285)$ sideband  and the green
cross-hatched histograms show the normalized continuum
contributions.} \label{f12data}
\end{figure*}

For $\Upsilon(1S,2S)\to f_1(1285)+X_{\rm tetra}$, $f_1(1285)$
candidates are reconstructed via $\eta\pi^+\pi^-$, $\eta\to \gamma
\gamma$. Figure~\ref{f12data} shows the recoil mass spectra of the
$f_1(1285)$ in $\Upsilon(1S,2S)$ data, together with the
backgrounds from the normalized $f_1(1285)$ sideband
and the normalized continuum contributions. No evident $X_{\rm
tetra}$ signals are seen. An unbinned extended maximum-likelihood
fit repeated with $X_{\rm tetra}$ masses from 1.46 to
2.46~GeV/$c^2$ in steps of 0.10~GeV/$c^2$, and with $X_{\rm
tetra}$ widths from 0.0 to 0.3~GeV in steps of 0.1~GeV, is applied
to the recoil mass spectra. The signal shape of each $X_{\rm
tetra}$ signal is described with a BW function
convolved with a Novosibirsk function, where all
parameter values are fixed to those from the fit to the
MC-simulated signals. Since no backgrounds showing peak distributions are found, a
second-order Chebyshev polynomial shape is used for the
backgrounds. The fit result for the $X_{\rm tetra}$ signal with
its mass fixed at 1.66~GeV/$c^2$ (a theoretically predicted mass for a scalar
tetraquark state~\cite{1610.02081}) and width fixed at 0.10~GeV is
shown in Fig.~\ref{f12data}. The fit yields $1.7\pm 4.7$ ($-0.3\pm
9.8$) events for the $X_{\rm tetra}$ signals in the $\Upsilon(1S)$
($\Upsilon(2S)$) data sample.
In the whole mass region of interest, the most significant signal is observed
at an $X_{\rm tetra}$ mass of 2.26 (2.16)~GeV/$c^2$ and width of 0.0 (0.3) GeV
with a statistical significance of 1.1$\sigma$ (1.8$\sigma$) in $\Upsilon(1S)$ $(\Upsilon(2S))$ data.

For $\chi_{b1} \to J/\psi+X_{\rm tetra}$, the $\chi_{b1}$ is
identified through the decay $\Upsilon(2S)\to \gamma \chi_{b1}$.
Figure~\ref{GBdata} shows the recoil mass spectrum of $\gamma
\jpsi$ in $\Upsilon(2S)$ data, together with the background
estimated from the normalized $\jpsi$ sideband and the
normalized continuum contributions. No evident $X_{\rm tetra}$
signal is observed. An unbinned extended maximum-likelihood fit is
applied to the $\gamma \jpsi$ recoil mass spectrum. The result of
the fit with the $X_{\rm tetra}$  mass fixed at
1.66~GeV/$c^2$ and width fixed at 0.10~GeV is shown in
Fig.~\ref{GBdata}. This fit yields $8.9\pm 5.8$ $X_{\rm tetra}$
signal events. In the entire region of study, the most significant signal is observed
at an $X_{\rm tetra}$ mass of 1.76~GeV/$c^2$ and width of 0.1 GeV,
with a statistical significance of 2.8$\sigma$.

\begin{figure}[htbp]
\includegraphics[height=7cm,angle=-90]{fig13.epsi}
\caption{(Color online) The $\gamma\jpsi$ recoil mass spectrum for
$\Upsilon(2S)\to \gamma \chi_{b1} \to \gamma \jpsi + anything$ in
the $\Upsilon(2S)$ data sample. The blue solid curve shows the result
of the fit described in the text, including the $X_{\rm tetra}$
state with a width fixed to 0.10~GeV and a mass fixed at
1.66~GeV/$c^2$ indicated by the arrow. The blue dashed curve shows the
fitted background. The magenta  shaded histogram is from the normalized
$\jpsi$  sideband  and the green cross-hatched histogram shows
the normalized continuum contributions.} \label{GBdata}
\end{figure}

\begin{figure}[htbp]
\includegraphics[height=7cm,angle=-90]{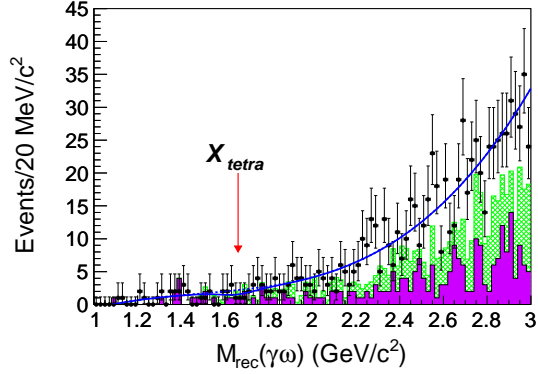}
\caption{(Color online) The $\gamma\omega$ recoil mass spectrum
for $\Upsilon(2S)\to \gamma \chi_{b1} \to \gamma \omega +
anything$ in the $\Upsilon(2S)$ data sample. The blue solid curve shows
the result of the fit described in the text, including the $X_{\rm
tetra}$ state with a width fixed to 0.10~GeV and a mass fixed at
1.66~GeV/$c^2$ indicated by the arrow. The blue dashed curve shows the
fitted background. The magenta  shaded histogram is from the normalized
$\omega$ sideband and the green cross-hatched histogram
shows the normalized continuum contributions.} \label{omegadata}
\end{figure}

For $\chi_{b1}\to \omega +X_{\rm tetra}$,  $\omega$ candidates are
reconstructed via $\pp\pi^0$, $\pi^0 \to \gamma \gamma$. Figure~\ref{omegadata} shows the
recoil mass spectrum of $\gamma \omega$ for events in the $\omega$
signal region, along with the backgrounds from the normalized $\omega$
sideband  and the normalized continuum
contributions. No evident $X_{\rm tetra}$ signal is observed. An
unbinned extended maximum-likelihood fit is applied to the $\gamma
\omega$ recoil mass spectrum. The result of the fit including the
$X_{\rm tetra}$ signal with its mass fixed at 1.66~GeV/$c^2$ and
width fixed at 0.10~GeV is shown in Fig.~\ref{omegadata}. This fit
yields $-7.8\pm 9.1$ $X_{\rm tetra}$ signal events.
In the entire region of study, the most significant signal is observed
at an $X_{\rm tetra}$ mass of 2.26~GeV/$c^2$ and width of 0.1 GeV,
with a statistical significance of 2.2$\sigma$.

Considering the yields for $\Upsilon(1S,2S)\to
f_1(1285)+X_{\rm tetra}$, $\chi_{b1}\to J/\psi+X_{\rm tetra}$ and
$\chi_{b1} \to \omega +X_{\rm tetra}$ are very small, we determine the 90\% C.L. upper limits
on the $X_{\rm tetra}$ signal yields ($N^{\rm UL}$) for $M(X_{\rm tetra})<1.46$~GeV/$c^2$
following the procedure in Ref.~\cite{D67.012002} as described
above for $\Upsilon(1S,2S)\to \chi_{c1}+X_{\rm tetra}$,
and for $M(X_{\rm tetra}) > 1.46\,{\rm GeV}/c^2$
using the same method as described for $\chi_{b0,b1}\to J/\psi+anything$.
Here, the systematic errors have been taken into account in the determination of $N^{UL}$.

\begin{figure*}[htbp]
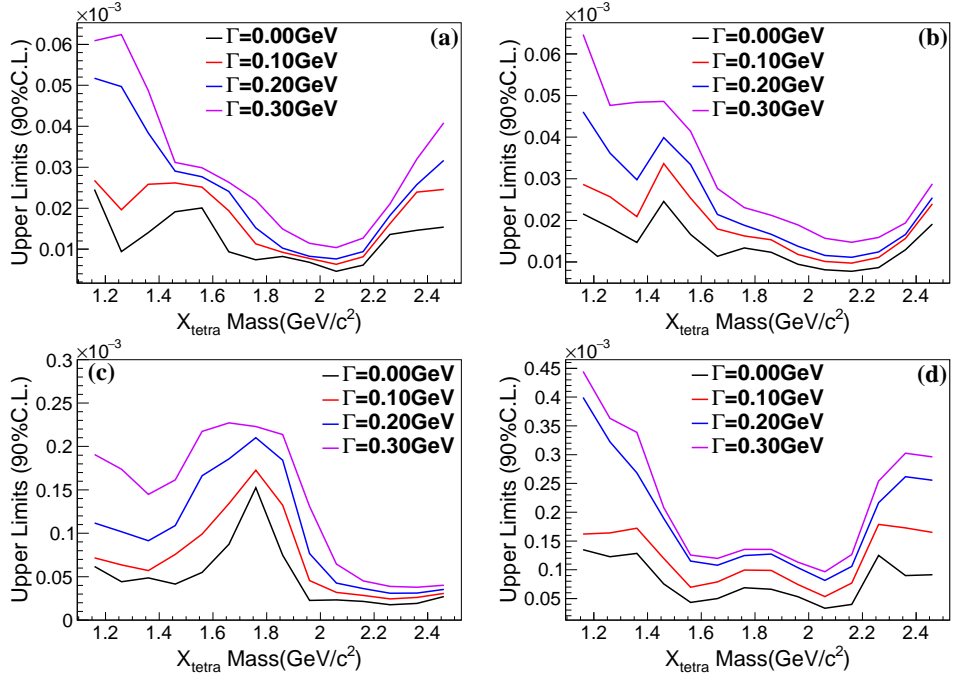

\includegraphics[height=6cm,angle=-90]{fig15a.epsi}
\hspace{0.3cm}
\includegraphics[height=6cm,angle=-90]{fig15b.epsi}
\hspace{0.3cm}
\includegraphics[height=6cm,angle=-90]{fig15c.epsi}
\hspace{0.3cm}
\includegraphics[height=6cm,angle=-90]{fig15d.epsi}
\caption{(Color online) The upper limits on the branching
fractions for (a) $\Upsilon(1S)\to f_1(1285)+X_{\rm tetra}$, (b)
$\Upsilon(2S) \to f_1(1285)+X_{\rm tetra}$, (c) $\chi_{b1} \to
\jpsi+X_{\rm tetra}$, and (d) $\chi_{b1} \to \omega+X_{\rm tetra}$
as a function of the assumed $X_{\rm tetra}$  mass with widths
fixed at 0.0, 0.1, 0.2, and 0.3~GeV, respectively.}\label{final2}
\end{figure*}

The calculated upper limits on the numbers of signal events
($N^{\rm UL}$) and branching fraction ($\BR^{\rm UL}$) for
$\Upsilon(1S,2S)\to f_1(1285)+X_{\rm tetra}$, $\chi_{b1}\to
J/\psi+X_{\rm tetra}$ and $\chi_{b1}\to \omega+X_{\rm tetra}$ with
$X_{\rm tetra}$ masses from 1.16 to 2.46~GeV/$c^2$ and widths from
0.0 to 0.3~GeV are listed in Table~\ref{Summary}, together with
the reconstruction efficiencies ($\varepsilon$) and the systematic
uncertainties ($\sigma_{\rm syst}$). The results are displayed
graphically in Fig.~\ref{final2}.

\section{\boldmath SYSTEMATIC Uncertainties}

Most of the systematic errors in the branching fraction
measurements are the same as in Ref.~\cite{oddball}, including
tracking reconstruction, photon reconstruction, particle identification,
trigger efficiency, the branching fractions of the
intermediate states, and the total numbers of $\Upsilon(1S)$
and $\Upsilon(2S)$ events; the notable exception is
the dominant systematic error from the fit uncertainty. By changing
the order of the background polynomial and the range of the fit,
the model-dependent relative difference in the
signal yields (or the upper limits  for those modes with
statistically insignificant branching fractions) is obtained; this is taken as the
systematic error due to the uncertainty of the fit.
The estimation of the continuum contributions in the $f_1(1285)$ inclusive production processes
assumes a  $1/s^{2}$ dependence. The analysis is repeated assuming a $1/s$ or $1/s^{3}$  dependence
and the largest change in the fitted $f_1(1285)$ signal yield is taken as a systematic uncertainty.
Assuming that all of these systematic-error sources are independent, the total
systematic errors are summed in quadrature and listed in
Table~\ref{Summary} for all the studied modes
for each hypothesized  $X_{\rm tetra}$ mass.

\section{\boldmath Summary}

In summary, utilizing the recoil mass spectra of the
$\chi_{c1}$, $f_1(1285)$, $\jpsi$, and $\omega$ in the channels
$\Upsilon(1S,2S)\to \chi_{c1}+G_{0^{--}}$, $\Upsilon(1S,2S)\to
f_1(1285)+G_{0^{--}}$, $\chi_{b1} \to J/\psi+G_{0^{--}}$, and
$\chi_{b1}\to \omega+G_{0^{--}}$~\cite{oddball}, respectively, we
report the first search for the light tetraquark states predicted
with a mass of $1.66\pm 0.14$~GeV/$c^2$ and $J^{PC}=0^{--}$,
and with a mass in the region 1.18--1.43 GeV/$c^2$
and $J^{PC}=1^{+-}$~\cite{1610.02081}. No evident signal is found
below 3~GeV/$c^2$ in the above processes and $90\%$
C.L. upper limits are set on the branching fractions.
Figures~\ref{final1} and~\ref{final2} show the upper limits on the
branching fractions as a function of the tetraquark masses.
In addition, as byproducts of the search, we measure the inclusive $f_1(1285)$
production in $\Upsilon(1S,2S)$, $J/\psi$ production in $\chi_{bJ} (J = 0, 1, 2)$, and $\omega$ production in $\chi_{b1}$.
The corresponding branching fractions are measured for the first time to be
$\BR(\Upsilon(1S)\to f_1(1285)+anything)=(46\pm 28({\rm stat.})\pm13({\rm syst.}))\times 10^{-4}$,
$\BR(\Upsilon(2S)\to f_1(1285)+anything)=(22\pm 15({\rm stat.})\pm6.3({\rm syst.}))\times 10^{-4}$, $\BR(\chi_{b2} \to J/\psi + anything)=(1.50\pm0.34({\rm stat.})\pm0.22({\rm syst.}))\times 10^{-3}$, and $\BR(\chi_{b1} \to \omega + anything)=(4.9\pm1.3({\rm stat.})\pm0.6({\rm syst.}))\times 10^{-2}$,
and the $90\%$ C.L. upper limits on the branching fractions $\BR(\chi_{b0} \to J/\psi + anything)<2.3\times 10^{-3}$ and
$\BR(\chi_{b1} \to J/\psi + anything)<1.1\times 10^{-3}$ are determined for the first time.

\begin{table*}[htbp]
\caption{\label{Summary} Summary of the upper limits for
$\Upsilon(1S,2S)\to \chi_{c1}+X_{\rm tetra}$, $f_1(1285)+X_{\rm
tetra}$, and $\chi_{b1}\to J/\psi+X_{\rm tetra}$, $\omega+X_{\rm
tetra}$ under different assumptions of $X_{\rm tetra}$ mass ($m$
in GeV/$c^2$) and width ($\Gamma$ in GeV), where $N^{\rm UL}$ is
the upper limit on the number of signal events taking into account
systematic errors, $\varepsilon$ is the reconstruction efficiency,
$\sigma_{\rm syst}$ is the total relative systematic uncertainty on the branching fraction and
$\BR^{\rm UL}$ is the 90\% C.L. upper limit on the branching
fraction. } \fontsize{6.2}{10}\selectfont
\begin{tabular}{c|cccc|cccc}
  \hline\hline
    \multicolumn{5}{c}{$\Upsilon(1S) \to \chi_{c1}+X_{\rm tetra}$ $({\rm for}~\Gamma=0.0/0.1/0.2/0.3$ GeV)} & \multicolumn{4}{c}{$\Upsilon(2S) \to \chi_{c1}+X_{\rm tetra}$ $({\rm for}~\Gamma=0.0/0.1/0.2/0.3$ GeV)} \\
    $m$ & $\varepsilon (\%)$ & $N^{\rm UL}$ &  $\sigma_{\rm syst} (\%)$ & $\BR^{\rm UL}(\times 10^{-6})$ & $\varepsilon (\%)$ & $N^{\rm UL}$ &  $\sigma_{\rm syst} (\%)$ & $\BR^{\rm UL}(\times 10^{-6})$ \\
  \hline
1.16&3.1/2.9/3.1/3.0&2.3                &6.2 &18.3/19.6/17.9/18.7&3.0/2.6/3.0/2.7    &4.7/4.7/4.7/6.0  &6.3 &26.2/29.1/25.7/36.7\\
1.26&4.4/4.2/4.5/4.3&2.3                &6.2 &12.7/13.3/12.5/12.9&4.3/4.0/4.2/4.1    &4.7/4.7/6.0/7.6  &6.3 &17.8/19.1/23.6/29.5\\
1.36&5.7/5.5/5.8/5.7&2.3                &6.2 &9.8/10.1/9.7/9.9   &5.7/5.4/5.6/5.5    &4.7/4.7/7.9/7.9  &6.3 &13.6/14.3/23.3/23.7\\
1.46&7.0/6.8/7.0/7.0&2.3/2.3/2.3/4.2    &6.2 &8.0/8.3/8.0/15.1   &6.7/6.5/7.0/6.9    &2.3/5.9/7.6/10.0  &6.3 &5.4/15.3/17.4/23.0\\
1.56&8.2/8.0/8.2/8.2&2.3/2.3/2.3/5.5    &6.2 &6.8/7.0/6.8/16.4    &8.2/8.0/8.1/8.1   &2.3/5.9/7.0/10.5  &6.3 &4.4/12.5/14.2/21.2\\
1.66&9.4/9.2/9.4/9.4&2.3/2.3/4.2/6.1    &6.2 &6.0/6.1/11.1/16.4    &9.2/9.0/9.3/9.3    &4.7/5.8/10.0/14.4  &6.3 &8.4/10.4/17.1/24.9\\
1.76&10.5/10.3/10.5/10.5&2.3/2.3/5.5/6.5&6.2 &5.3/5.4/12.7/15.5    &10.3/10.1/10.4/10.3&5.8/6.7/8.8/14.9   &6.3 &9.0/10.7/13.5/23.0\\
1.86&11.6/11.4/11.6/11.6&2.3/4.2/6.1/7.1&6.2 &4.8/9.2/13.2/15.7    &11.4/11.2/11.3/11.4&6.7/8.7/12.1/17.8  &6.3 &9.5/12.2/18.0/25.3\\
1.96&12.7/12.7/12.5/12.7&2.3/5.5/6.5/9.2&6.2 &4.4/10.6/13.0/17.8    &12.5/12.5/12.4/12.5&4.2/9.3/13.5/17.3 &6.3 &5.4/11.6/17.4/22.8\\
2.06&13.7/13.5/13.7/13.7&4.1/6.1/7.1/10.2&6.2 &8.7/11.3/13.3/19.3   &13.6/13.4/13.5/13.5&6.7/11.2/16.7/21.1 &6.3 &8.0/13.3/19.9/26.8\\
2.16&14.7/14.5/14.6/14.7&5.5/7.2/9.2/12.9&6.2 &9.1/12.9/15.4/22.3  &14.6/14.4/14.5/14.5&4.7/10.5/17.3/24.6  &6.3 &5.3/11.9/19.8/28.8\\
2.26&15.6/15.4/15.6/15.7&6.5/7.3/10.2/16.4&6.2 &10.7/12.4/17.0/25.9 &15.5/15.3/15.5/15.4&10.1/13.5/20.7/30.4&6.3 &10.4/14.1/22.0/33.1\\
2.36&16.6/16.4/16.5/16.6&4.1/9.3/13.8/20.7&6.2 &7.2/13.8/20.8/32.4  &16.4/16.2/16.4/16.4&10.1/14.3/29.3/30.0&6.3 &9.9/14.8/28.3/30.6\\
2.46&17.4/17.2/17.4/17.5&4.1/9.3/16.9/24.3&6.2&6.8/13.2/24.0/37.8  &17.3/17.1/17.3/17.3&10.4/18.7/27.7/32.3&6.3 &10.1/18.1/26.3/31.2\\
  \hline\hline

  \multicolumn{5}{c}{$\Upsilon(1S) \to f_1(1285)+X_{\rm tetra}$ $({\rm for}~\Gamma=0.0/0.1/0.2/0.3$ GeV)} & \multicolumn{4}{c}{$\Upsilon(2S) \to f_1(1285)+X_{\rm tetra}$
  $({\rm for}~\Gamma=0.0/0.1/0.2/0.3$ GeV)} \\
    $m$ & $\varepsilon (\%)$ & $N^{\rm UL}$ &  $\sigma_{\rm syst} (\%)$ & $\BR^{\rm UL}(\times 10^{-6})$ & $\varepsilon (\%)$ & $N^{\rm UL}$ &  $\sigma_{\rm syst} (\%)$ & $\BR^{\rm UL}(\times 10^{-6})$ \\
  \hline
1.16&1.2/1.1/1.0/1.0&4.4/4.4/7.8/9.1    &20.2               &24.5/26.8/51.7/60.9&1.0/0.9/1.0/1.0&5.1/6.0/10.8/15.1  &20.2&21.6/28.6/46.1/64.7\\
1.26&1.6/1.5/1.7/1.5&2.3/4.4/12.7/14.1  &20.2               &9.4/19.6/49.7/62.4&1.5/1.4/1.6/1.5&6.4/8.4/13.5/16.6  &20.2 &18.3/25.7/36.2/47.6\\
1.36&2.1/2.0/2.2/2.0&4.4/7.8/12.7/14.8  &20.2               &14.0/25.9/38.4/48.8&2.1/2.0/2.2/2.0&7.2/9.8/15.3/22.7 &20.2 &14.7/20.9/29.8/48.4\\
1.46&2.6/2.7/2.5/2.5&7.6/10.7/11.0/11.6 &20.9/21.6/22.2/24.4&19.1/26.2/29.1/31.2&2.6/2.5/2.5/2.4&15.0/19.7/23.3/27.3&20.4/24.9/27.8/28.6 &24.5/33.6/39.9/48.6\\
1.56&3.2/3.2/3.1/3.0&9.7/12.1/12.9/13.4 &20.2/20.2/21.5/22.2&20.0/25.2/27.7/29.9&3.0/3.0/3.1/3.0&12.1/17.9/24.2/28.9&20.1/20.4/22.3/24.4 &16.6/25.4/33.4/41.4\\
1.66&3.6/3.5/3.7/3.5&5.2/10.3/13.4/13.9 &20.5/21.7/22.0/24.5&9.4/19.4/24.1/26.3 &3.5/3.4/3.6/3.5&9.3/14.3/18.1/22.5&22.0/23.2/23.5/28.3 &11.4/17.9/21.4/27.7 \\
1.76&4.1/4.0/4.2/4.0&4.5/6.8/9.8/13.3   &20.5/21.6/24.2/24.3&7.4/11.3/15.2/21.9 &4.0/3.9/4.1/4.0&12.5/14.9/18.2/21.4&20.8/23.6/24.5/29.1&13.4/16.3/18.8/23.1 \\
1.86&4.5/4.4/4.6/4.4&5.5/6.6/7.1/9.9   &20.3/20.5/21.0/22.3&8.2/9.2/10.2/14.9  &4.4/4.3/4.5/4.4&12.7/15.4/17.4/21.7&20.5/21.4/21.5/23.0&12.8/15.3/16.7/21.2 \\
1.96&4.9/4.8/5.0/4.8&5.1/5.8/6.2/8.3    &20.3/20.3/20.6/21.0&6.8/7.7/8.2/11.4   &4.8/4.7/4.9/4.7&10.6/12.9/15.9/20.8&20.2/20.6/27.0/23.7&9.4/11.8/13.8/18.9 \\
2.06&5.3/5.2/5.4/5.2&3.7/5.0/6.2/8.1    &20.0/20.2/20.2/20.3&4.6/6.3/7.6/10.4   &5.2/5.1/5.3/5.1&9.9/12.0/14.3/18.8&24.8/26.0/27.0/27.5&8.1/10.1/11.6/15.7 \\
2.16&5.7/5.6/5.8/5.6&5.3/6.9/8.2/10.8   &20.4/21.7/23.7/24.2&6.1/8.2/9.4/12.7   &5.6/5.5/5.7/5.5&10.2/12.6/14.8/19.0&26.7/27.4/30.8/33.6&7.8/9.7/11.1/14.7 \\
2.26&6.1/6.0/6.2/6.0&12.6/14.7/17.0/19.0&21.4/24.0/24.2/30.9&13.6/16.2/18.2/21.0&6.0/5.9/6.1/5.9&12.1/15.3/17.8/22.0&21.5/24.8/28.8/35.2&8.6/11.1/12.4/15.9 \\
2.36&6.5/6.4/6.6/6.4&14.2/22.8/25.7/30.7&24.5/27.7/28.8/32.9&14.6/23.9/25.8/32.0&6.4/6.3/6.5/6.4&19.4/23.0/25.3/28.6&20.6/25.6/26.5/28.0&12.9/15.7/16.6/19.4 \\
2.46&6.8/6.7/6.9/6.7&15.9/24.8/32.4/40.1&20.6/21.1/21.2/22.3&15.4/24.6/31.7/40.8&6.7/6.6/6.8/6.6&29.7/36.8/40.1/42.7&20.9/22.2/23.5/29.9&19.1/23.9/25.4/28.7 \\
  \hline\hline

  \multicolumn{5}{c}{$\chi_{b1} \to \jpsi+X_{\rm tetra}$ $({\rm for}~\Gamma=0.0/0.1/0.2/0.3$ GeV)} & \multicolumn{4}{c}{$\chi_{b1} \to \omega+X_{\rm tetra}$ $({\rm for}~\Gamma=0.0/0.1/0.2/0.3$ GeV)}
  \\
    $m$ & $\varepsilon (\%)$ & $N^{\rm UL}$ &  $\sigma_{\rm syst} (\%)$ & $\BR^{\rm UL}(\times 10^{-5})$ & $\varepsilon (\%)$ & $N^{\rm UL}$ &  $\sigma_{\rm syst} (\%)$ & $\BR^{\rm UL}(\times 10^{-5})$ \\
  \hline
1.16&2.4/2.5/2.3/2.3&1.9/2.3/3.3/5.6    &7.8  &6.2/7.2/11.8/19.1&0.4/0.5/0.4/0.6  &5.7/7.2/15.3/24.0  &9.3 &13.5/16.2/40.0/44.4\\
1.26&3.7/3.8/3.6/3.6&2.1/3.1/4.6/8.0    &7.8  &4.4/6.4/10.2/17.4&0.7/0.7/0.7/0.7  &8.6/11.6/21.1/25.6 &9.3 &12.2/16.4/32.2/26.3\\
1.36&4.9/5.0/4.8/4.9&3.1/3.7/5.6/9.0   &7.8  &4.9/5.7/9.1/14.5&1.0/1.0/1.0/1.1    &12.6/16.5/24.6/34.4 &9.3 &12.8/17.2/26.8/33.9\\
1.46&6.1/6.2/6.0/6.1&3.3/6.0/8.4/12.5   &7.9/8.7/10.1/13.9&4.2/7.6/10.9/16.2&1.3/1.3/1.2/1.4&9.4/14.5/22.3/27.1&19.0/20.3/21.4/23.5&7.5/12.0/18.9/20.8\\
1.56&7.3/7.4/7.2/7.3&5.2/9.4/15.3/20.2  &7.9/8.1/8.3/9.1&5.5/9.9/16.7/21.7&1.6/1.5/1.5/1.6&6.6/10.1/16.4/19.3  &11.4/14.0/18.6/19.8&4.3/7.0/11.4/12.6\\
1.66&8.4/8.5/8.3/8.4&9.4/14.6/19.8/24.4 &7.9/8.0/11.4/12.4&8.7/13.4/18.6/22.7 &1.9/1.8/1.8/1.9&8.8/13.2/18.0/21.2  &13.1/13.9/16.4/16.9&5.0/7.9/10.8/12.0\\
1.76&9.5/9.6/9.4/9.5&18.6/21.3/25.4/27.1&8.1/9.0/9.9/12.0 &15.2/17.2/21.0/22.3&2.1/2.0/2.0/2.1&13.9/19.0/23.8/27.1 &9.5/10.9/13.4/13.7 &6.9/9.9/12.4/13.5\\
1.86&10.6/10.7/10.5/10.6&10.2/18.2/24.8/28.7&7.9/8.6/8.9/10.4 &7.5/13.2/18.4/21.4 &2.4/2.2/2.2/2.4&15.0/21.3/27.3/30.2 &10.0/11.0/11.1/11.3&6.6/9.9/12.7/13.5\\
1.96&11.6/11.7/11.5/11.6&3.4/6.8/11.2/19.5  &8.0/8.1/8.9/11.1 &2.3/4.6/7.7/13.1   &2.6/2.5/2.5/2.6&13.2/17.5/24.4/27.5 &9.3/9.4/9.5/10.7   &5.3/7.4/10.3/11.2\\
2.06&12.6/12.7/12.5/12.6&3.8/5.2/6.8/10.4   &7.9/8.0/8.1/8.3  &2.3/3.2/4.3/6.4   &2.8/2.7/2.7/2.8&9.0/13.7/21.2/25.7  &9.7/9.9/10.1/10.4  &3.3/5.3/8.1/9.6\\
2.16&13.6/13.7/13.5/13.6&3.8/5.0/6.2/7.8    &7.8/8.1/8.2/8.4  &2.2/2.8/3.6/4.5   &3.1/3.0/3.0/3.0&11.7/21.3/29.8/36.1 &10.4/11.4/11.6/12.0&4.0/7.7/10.5/12.6\\
2.26&14.5/14.6/14.4/14.5&3.3/4.6/5.7/7.2    &7.9/8.1/8.6/10.8 &1.8/2.4/3.1/3.9  &3.3/3.1/3.2/3.2&39.1/52.9/64.9/76.7 &10.0/10.3/10.7/12.8&12.5/17.9/21.6/25.4\\
2.36&15.4/15.5/15.3/15.4&3.8/5.2/6.1/7.5    &7.9/9.2/9.5/12.9 &1.9/2.6/3.1/3.8   &3.5/3.3/3.4/3.4&30.2/54.9/84.4/96.7 &13.0/14.3/15.2/16.2&9.0/17.3/26.2/30.2\\
2.46&16.2/16.3/16.1/16.2&5.7/6.4/7.3/8.3    &8.2/8.5/8.6/8.7  &2.7/3.1/3.5/4.0   &3.7/3.5/3.6/3.6&32.4/55.7/86.8/100.9&15.0/15.3/17.3/18.0&9.1/16.5/25.5/29.6\\
  \hline\hline
  \end{tabular}
\end{table*}

\section{ACKNOWLEDGMENTS}
We thank the KEKB group for the excellent operation of the
accelerator; the KEK cryogenics group for the efficient
operation of the solenoid; and the KEK computer group,
the National Institute of Informatics, and the
PNNL/EMSL computing group for valuable computing
and SINET5 network support.  We acknowledge support from
the Ministry of Education, Culture, Sports, Science, and
Technology (MEXT) of Japan, the Japan Society for the
Promotion of Science (JSPS), and the Tau-Lepton Physics
Research Center of Nagoya University;
the Australian Research Council;
Austrian Science Fund under Grant No.~P 26794-N20;
the National Natural Science Foundation of China under Contracts
No.~10575109, No.~10775142, No.~10875115, No.~11175187, No.~11475187,
No.~11521505 and No.~11575017;
the Chinese Academy of Science Center for Excellence in Particle Physics;
the Ministry of Education, Youth and Sports of the Czech
Republic under Contract No.~LTT17020;
the Carl Zeiss Foundation, the Deutsche Forschungsgemeinschaft, the
Excellence Cluster Universe, and the VolkswagenStiftung;
the Department of Science and Technology of India;
the Istituto Nazionale di Fisica Nucleare of Italy;
the WCU program of the Ministry of Education, National Research Foundation (NRF)
of Korea Grants No.~2011-0029457, No.~2012-0008143,
No.~2014R1A2A2A01005286,
No.~2014R1A2A2A01002734, No.~2015R1A2A2A01003280,
No.~2015H1A2A1033649, No.~2016R1D1A1B01010135, No.~2016K1A3A7A09005603, No.~2016K1A3A7A09005604, No.~2016R1D1A1B02012900,
No.~2016K1A3A7A09005606, No.~NRF-2013K1A3A7A06056592;
the Brain Korea 21-Plus program, Radiation Science Research Institute, Foreign Large-size Research Facility Application Supporting project and the Global Science Experimental Data Hub Center of the Korea Institute of Science and Technology Information;
the Polish Ministry of Science and Higher Education and
the National Science Center;
the Ministry of Education and Science of the Russian Federation and
the Russian Foundation for Basic Research;
the Slovenian Research Agency;
Ikerbasque, Basque Foundation for Science and
MINECO (Juan de la Cierva), Spain;
the Swiss National Science Foundation;
the Ministry of Education and the Ministry of Science and Technology of Taiwan;
and the U.S.\ Department of Energy and the National Science Foundation.


\begin{thebibliography}{**}
\bibitem{Nov} The Novosibirsk function is defined as $f(x) =
\exp[-\frac12(\ln^2(1+\Lambda(x-x_0))/\tau^2+\tau^2)]$ with
$\Lambda=$ $\sinh(\tau\sqrt{\ln 4})/(\sigma\sqrt{\ln4})$. The
parameters represent the mean ($x_0$), the width ($\sigma$) and
the tail asymmetry ($\tau$).
\end{thebibliography}
\end{document}